\documentclass{icrc29}
\usepackage{graphicx,amssymb,amsmath,times}
\begin{document}
\title[Simulation of a Hybrid Extenstion to IceCube ...]{Simulation of a Hybrid Optical/Radio/Acoustic Extension to IceCube for EeV Neutrino Detection}

\author[D. Besson et al.] { D. Besson$^a$, S. B\"{o}ser$^b$, R. Nahnhauer$^b$, P.B. Price$^c$, and
	\newauthor
	J. A. Vandenbroucke$^c$ (justin@amanda.berkeley.edu) for the IceCube Collaboration$^d$\\
        (a) Dept. of Physics and Astronomy, University of Kansas, Lawrence,
            KS 66045-2151, USA \\
        (b) DESY, D-15738 Zeuthen, Germany \\
        (c) Dept. of Physics, University of California, Berkeley, CA 94720, USA \\
	(d) For a full author list, see arXiv:astro-ph/0509330
        }
\presenter{Presenter: R. Nahnhauer, ger-nahnhauer-R-abs1-og25-oral}

\maketitle

\begin{abstract}

Astrophysical neutrinos at $\sim$EeV energies promise to be an interesting source for 
astrophysics and particle physics. Detecting the predicted 
cosmogenic (``GZK'') neutrinos at 10$^{16}$ - 10$^{20}$ eV would test models of 
cosmic ray production at these energies and probe particle physics at $\sim$100~TeV
center-of-mass energy. While IceCube could detect $\sim$1 GZK event per year, it 
is necessary to detect 10 or more events per year in order to study temporal, 
angular, and spectral distributions. The IceCube observatory may be able to 
achieve such event rates with an extension including optical, radio, and 
acoustic receivers.  We present results from simulating such a hybrid detector. 

\end{abstract}

\section{Introduction}
Detecting and characterizing astrophysical neutrinos in the 10$^{16}$~eV to 10$^{20}$~eV range is a central 
goal of astro-particle
physics.  The more optimistic flux models in this range involve discovery physics including topological defects and
relic neutrinos.  Detecting the smaller flux of cosmogenic 
(or Greisen, Zatsepin, and Kusmin, ``GZK'')
neutrinos produced via ultra-high energy cosmic ray interaction with the cosmic microwave background 
 would test models
of cosmic ray production and propagation and of particle physics at extreme energies.
  With $\sim$100 detected events, their angular distribution
would give a measurement of the total neutrino-nucleon cross section at $\sim$100 TeV center of mass,
probing an energy scale well beyond the reach of the LHC.  Hence, as a baseline, a detector capable 
of detecting $\sim$10 GZK events per year
has promising basic physics potential.  If any of the more exotic theories predicting greater EeV neutrino
fluxes is correct,
the argument in favor of such a detector is even stronger. 

To detect $\sim$10 GZK events per year,
a detector with an effective volume of $\sim$100~km$^3$ at EeV energies 
is necessary.
In addition to the possibility of identifying neutrino-induced air showers, there are three 
methods of ultra-high energy neutrino detection in solid media: optical, 
radio, and acoustic.  Optical Cherenkov detection is a well-established 
technique that has detected 
atmospheric neutrinos up to 10$^{14}$ eV and set limits up to
10$^{18}$~eV \cite{Chirkin}.  Radio efforts 
have produced steadily improving upper limits on neutrino fluxes from 10$^{16}$ to
10$^{25}$~eV \cite{radio}. Acoustic detection efforts are at an earlier stage, with one limit published thus far 
from 10$^{22}$ to 10$^{25}$~eV
 \cite{Vandenbroucke05}.

The currently planned 1~km$^3$ 
optical neutrino telescopes expect a GZK event rate of $\sim$1 per year.
It is possible to extend this by adding more optical strings for a modest
additional cost \cite{Halzen}, but it's difficult to imagine achieving 10 or more events per year with
optical strings alone.  The radio and acoustic methods have potentially large effective volumes with relatively
few receivers, but the methods are unproven in that they have never detected a neutrino.  Indeed, if
radio experiments claim detection of a GZK signal, it may be difficult to confirm that it is really a
neutrino signal.
However, it may be possible to bootstrap the large effective volumes of radio and acoustic detection
with the optical method, by building a hybrid detector that can detect a large
rate of radio or acoustic events, a fraction of which are also detected by an optical detector.  
A signal seen
in coincidence between any two of the three methods
would be convincing. 
The information from multiple methods can be combined for hybrid reconstruction, yielding improved
angular and energy resolution.

We simulated the sensitivity of a detector that could be constructed by expanding the IceCube observatory 
currently under construction 
at the South Pole.  The ice at the South Pole is likely well-suited for all three methods:  Its optical clarity
has been established by the AMANDA experiment \cite{Chirkin}, and its radio clarity and suitability for radio detection in the 
GZK energy range has been established by the RICE experiment \cite{radio}.  
Acoustically, the signal in ice is ten times greater than that in water.
Theoretical estimates indicate 
low attenuation and noise
\cite{Price05}, and efforts are planned to measure both \cite{SPATS05} 
with sensitive transducers
developed for glacial ice \cite{Nahnhauer05}.
Here we estimate the sensitivity of such a detector by exposing all three components to a common
Monte Carlo event set and counting events
detected by each method alone and by each combination of multiple methods.

\section{Simulation}

IceCube will have 80 strings arrayed hexagonally with a horizontal spacing of 125~m.  In \cite{Halzen},
the GZK sensitivity achieved by adding more optical strings at larger distances (``IceCube-Plus'')
was estimated, and the possibility of also adding radio and acoustic modules was mentioned. 
Here we consider an IceCube-Plus configuration consisting of 
a ``small'' optical array overlapped by a ``large'' acoustic/radio array
with a similar number of strings but larger horizontal spacing.  
The optimal string spacing for GZK detection was found to be $\sim$1~km
for both radio and acoustic strings.
This coincidence allows the two methods to share hole drilling and cable
costs, both of which are dominant costs of such arrays.

\begin{figure}
\begin{minipage}[t]{7cm}
\includegraphics[width=1.0\linewidth]{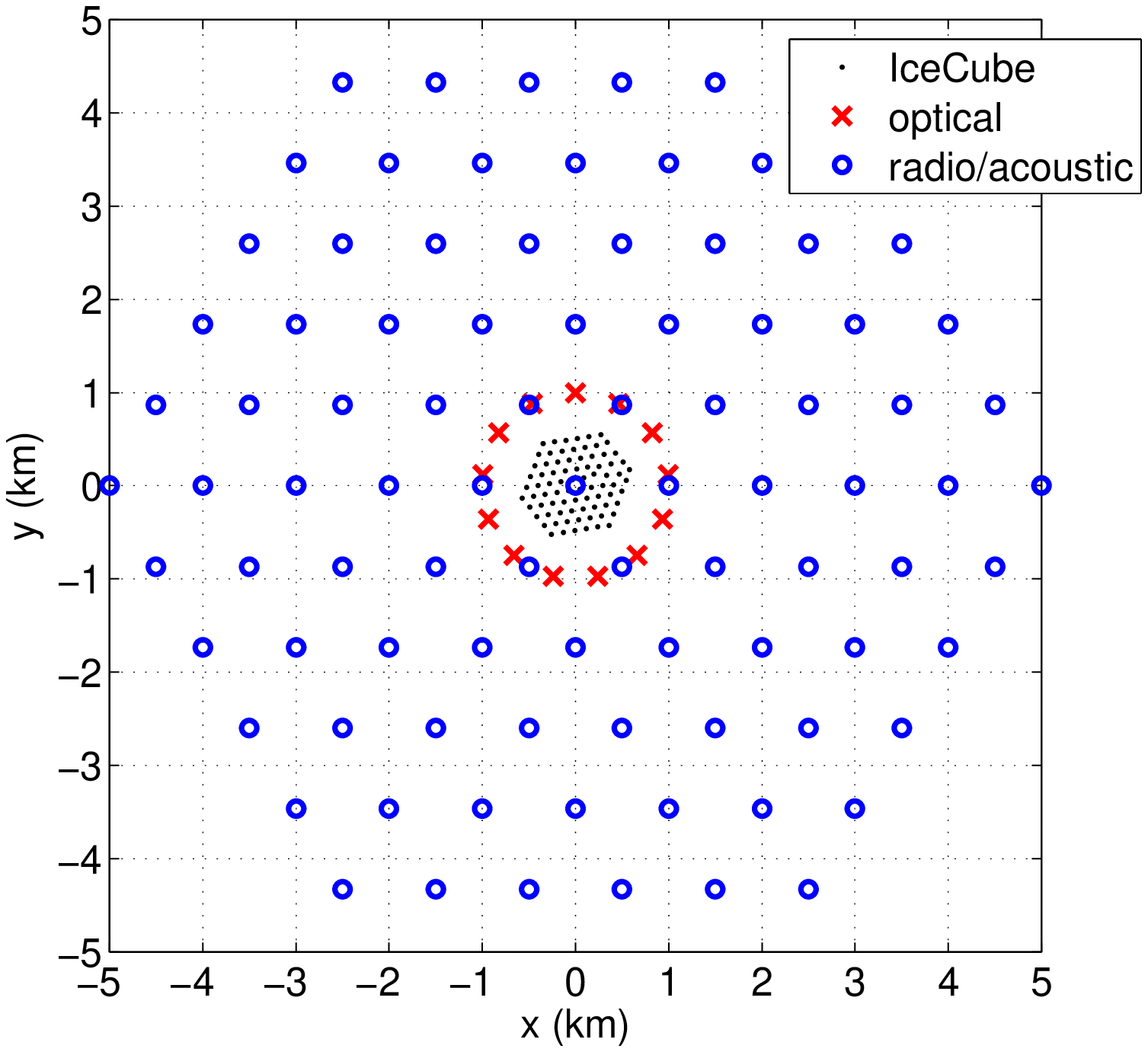}
\caption{\label {fig1} Geometry of the simulated hybrid array.}
\end{minipage}
\hfill
\begin{minipage}[t]{8.5cm}
\includegraphics[width=1.0\linewidth]{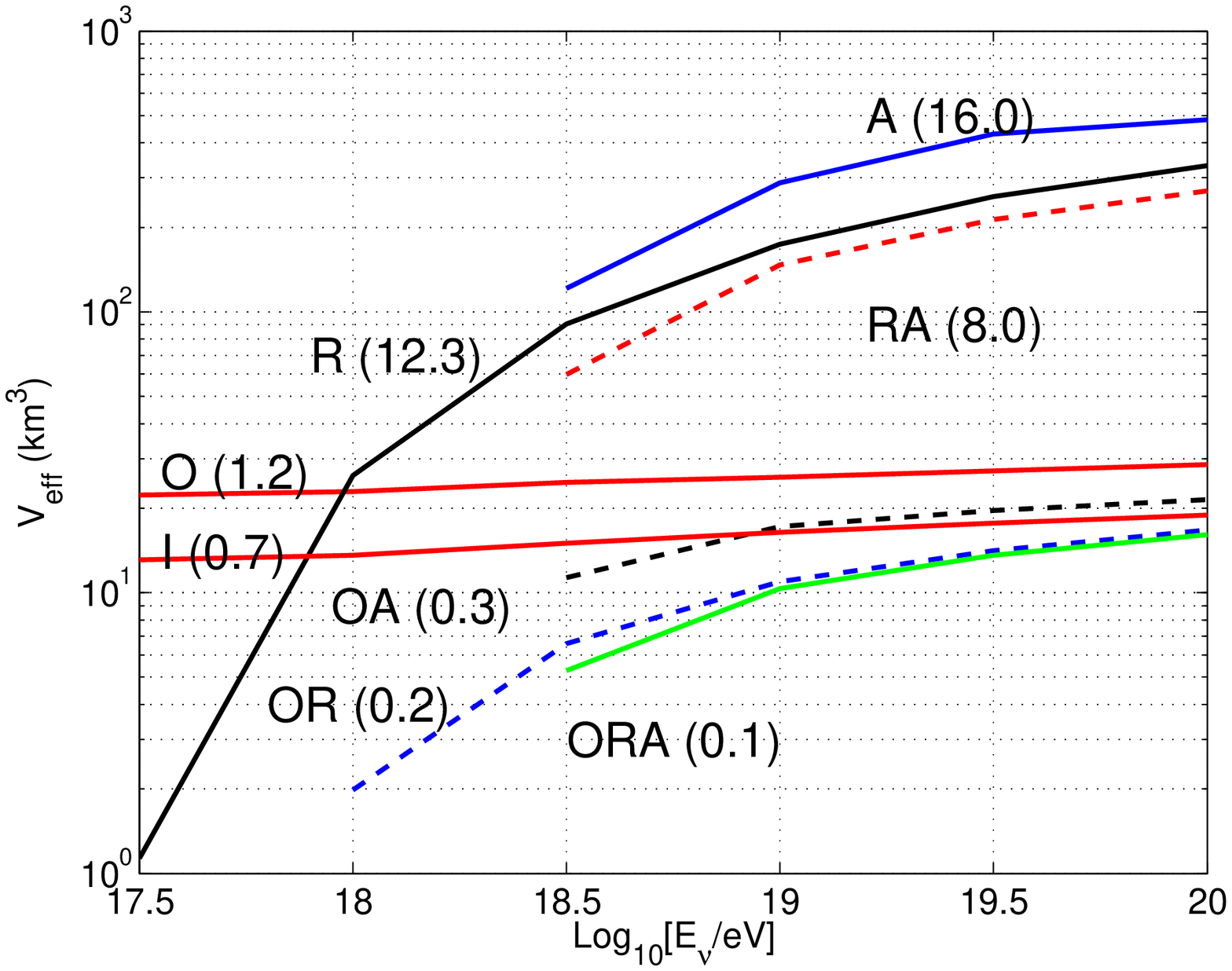}
\caption{\label {fig2} Effective volume for each of the seven
combinations of detector components, as well as for IceCube alone (``I'').
GZK event rates per year are given in parenthesis.  Note that different channels
were used for different combinations (see text).}
\end{minipage}
\hfill
\end{figure}

The geometry of the simulated array is shown in Fig. \ref{fig1}.
We take the optical array to be IceCube as well as a ring
of 13 optical strings with a 1~km radius, surrounding IceCube.
All optical strings have standard IceCube geometry: 60 modules per
string, spaced every 17 m, from 1.4 to 2.4~km depth.
Encompassing this is a hexagonal array of 91 radio/acoustic strings
with 1~km spacing.  Each radio/acoustic hole has 5 radio receivers, spaced every 100~m from 
200~m to 600~m depth, and 300 acoustic receivers, spaced every 5~m from 5~m to 1500~m depth.  
At greater depths both methods suffer increased absorption due to the warmer ice.  The 
large acoustic density per string is necessary because the acoustic
radiation pattern is thin (only $\sim$10~m thick) in the direction along the shower.  
The array geometry was designed to seek
an event rate of $\sim$10 GZK events per year detectable with both radio and acoustic
independently.
To obtain rough event rate estimates, a very simple Monte Carlo generation scheme was chosen.  
Between 10$^{16}$ and 10$^{20}$~eV, the neutrino interaction length ranges between 6000 and 200~km
\cite{Gandhi},
so upgoing
neutrinos are efficiently absorbed by the Earth and only downgoing events are detectable.  A
full simulation would include the energy-dependent slow rolloff at the horizon.  Here we
assume all upgoing neutrinos are absorbed before reaching the fiducial volume, and no
downgoing neutrinos are; we generate incident neutrino directions isotropically in 2$\pi$~sr.
Vertices are also generated uniformly in a fiducial cylinder of radius 10 km, extending
from the surface to 3 km depth.  

The Bjorken parameter $y = E_{had}/E_{\nu}$ varies
somewhat with energy and from event to event, but we choose the mean value, $y = $ 0.2, for simplicity.
The optical method can detect both muons and showers, but here we only
consider the muon channel; simulation of the shower channel is in progress.  
The radio and acoustic methods cannot detect muon tracks but
can detect electromagnetic and hadronic showers. Under our assumptions of constant $y$
and no event-to-event fluctuations, 
all flavors interacting via both CC and NC produce the same hadronic shower.  Electron
neutrinos interacting via the charged current also produce  an 
electromagnetic cascade which produces radio and acoustic
signals superposed on the hadronic signals.  However, at the energies of interest here,
electromagnetic showers are lengthened to hundreds of meters by the Landau-Pomeranchuk-Migdal effect.  
This weakens their radio and acoustic signals significantly, and we assume
they are negligible.

For simulation of the optical response, the standard Monte Carlo chain 
used in current AMANDA-IceCube analyses \cite{Chirkin} was performed.
After the primary trigger requiring any 5 hits in a 2.5~$\mu$s 
window, a
local coincidence trigger was applied: Ten local coincidences were required, where
a local coincidence is at least two hits on neighboring or next-to-neighboring 
modules within 1~$\mu$s. Compared with \cite{Halzen}, we used an updated ice
model with increased absorption, which may account for our factor of $\sim$2 lower effective
volume.

Each simulated radio ``receiver'' consists of two vertical
half-wave dipole antennas separated vertically by 5~m to allow local rejection
of down-going anthropogenic noise. We assume an effective height at the
peak frequency (280~MHz in ice) equal to 10~cm, with
$\pm$20\% bandwidth to the --3 dB points. As currently under development
for RICE-II, we assume optical fiber transport of the signal to the
DAQ, with losses of 1~dB/km (measured) through the fiber.
The electric field strength $E(\omega)$ is calculated
from the shower according to the ZHS prescription \cite{ZHS91,Alvarez-Muniz}.  Frequency-dependent 
ice attenuation effects are incorporated using measurements at
South Pole Station \cite{Barwick04}. The signal at the surface electronics 
is then transformed into the time domain, resulting in a waveform 10~ns long,
sampled at 0.5~ns intervals, at each antenna. Two
receivers with signals exceeding 3.5 times the
estimated rms noise temperature $\sigma_{kT}$ (thermal plus a system
temperature of 100~K) within a time window of 30~$\mu$s are required to trigger. 

The unattenuated acoustic pulse $P(t)$ produced 
at arbitrary position with
respect to a hadronic cascade is calculated by integrating over the cascade energy distribution.
The cascade is parametrized with the Nishimura-Kamata-Greisen parametrization, with 
$\lambda$ (longitudinal
tail length) parametrized from \cite{Alvarez-Muniz}.
The dominant mechanism of acoustic wave absorption in South Pole ice is theorized \cite{Price05} to be 
molecular reorientation, which increases with ice temperature.  Using a temperature profile 
measured at the South Pole
along with laboratory absorption measurements, an absorption vs. depth profile was estimated.
The predicted absorption length ranges from 8.6~km at the surface to 4.8~km at 1~km depth to 0.7~km at 
2~km depth.
The frequency-independent absorption is integrated from source to receiver and applied in the time domain.

South Pole ice is predicted to be much quieter than
ocean water at the relevant frequencies ($\sim$10-60 kHz), because there are no waves, currents, or animals.  Anthropogenic
surface noise will largely be waveguided back up to the surface due to the sound speed gradient
in the upper 200 m of uncompactified snow (``firn'').
For the current simulation we assume ambient noise
is negligible compared to transducer self-noise.  
Work is underway to produce transducers with self-noise at the 2-5~mPa level \cite{Nahnhauer05}. 
For comparison, ambient noise in the ocean is $\sim$100 mPa \cite{Vandenbroucke05}.
The acoustic trigger used in this simulation required that 3
receivers detect pressure pulses above a threshold of 9~mPa. 

\section{Results and Conclusion}

Ten-thousand events were generated at each half-decade in neutrino energy 
in a cylinder of volume 942~km$^3$.  For each
method and combination of methods, the number of detected events was used to
calculate effective volume as a function of neutrino energy (Fig. \ref{fig2}).  This was folded with the
GZK flux model of \cite{ESS,Seckel} and the cross-section
parametrizations of \cite{Gandhi} to estimate detectable
event rates (Fig. \ref{fig2}).  We use a flux model which assumes source evolution
according to $\Omega_{\Lambda}=$~0.7.  This model is a factor of $\sim$2
greater than that for $\Omega_{\Lambda}=$~0 evolution; it is unclear which model is
correct \cite{Seckel}. For radio and acoustic, and their combination,
all flavors and both interactions were included.  For those combinations including
the optical method, only the muon channel has been simulated thus far; including
showers will increase event rates for these combinations.  

It may be possible to build an extension like that considered here for a relatively small cost.
Holes for radio antennas and acoustic transducers can be narrow and shallow, and both
devices are simpler than photo-multiplier tubes.
The necessarily large acoustic channel multiplicity is partially offset by the fact that the 
acoustic signals are slower by five orders of magnitude, making data acquisition and processing easier.

The IceCube observatory will observe the neutrino universe from 10's of 
GeV to 100's of PeV.
Our simulations indicate that extending it with radio and acoustic strings could produce a neutrino detector
competitive with other projects optimized for high-statistics measurements of GZK neutrinos but with the unique 
advantage of cross-calibration via
coincident optical-radio, optical-acoustic, and radio-acoustic events.

\end{document}